# Multimodal Analysis of Traction Forces and Temperature Dynamics of Living Cells with Diamond-Embedded Substrate

Tomasz Kołodziej * 1),2),, Mariusz Mrózek 2), Saravanan Sengottuvel 2), Maciej J. Głowacki 3), Mateusz Ficek 3), Wojciech Gawlik 2), Zenon Rajfur 2), Adam Wojciechowski * 2)

*1) Jagiellonian University Medical School, Faculty of Pharmacy, Kraków, Poland*

*2) Jagiellonian University, Faculty of Physics, Astronomy and Applied Computer Science, Kraków, Poland*

*3) Gdansk University of Technology, Faculty of Electronics, Telecommunications and Informatics, Department of Metrology and Optoelectronics, Gdańsk, Poland*

Corresponding authors: Tomasz Kołodziej, tomasz.kolodziej@uj.edu.pl, Adam Wojciechowski, adam.wojciechowski@uj.edu.pl

**Abstract**

Cells and tissues are constantly exposed to various chemical and physical signals that intricately regulate various physiological and pathological processes. This study explores the integration of two biophysical methods, Traction Force Microscopy (TFM) and Optically-Detected Magnetic Resonance (ODMR), to concurrently assess cellular traction forces and local relative temperature. We present a novel elastic substrate with embedded nitrogen-vacancy microdiamonds, that facilitate ODMR-TFM measurements. Optimization efforts have focused on minimizing the sample illumination and experiment duration to mitigate biological perturbations. Our hybrid ODMR-TFM technique yields precise TFM maps and achieves approximately 1K accuracy in relative temperature measurements. Notably, our setup, employing a simple wide-field fluorescence microscope with standard components, demonstrates the broader feasibility of these techniques in life-science laboratories. By elucidating the physical aspects of cellular behavior beyond the existing methods, this approach opens avenues for a deeper understanding and may inspire diverse biomedical applications.

## Introduction

Cells and tissues are complex biological objects regulated by a variety of chemical and physical signals. Chemical stimulation is induced by sugars, fatty acids, amino acids, mineral salts, vitamins, hormones, and other chemical compounds that are necessary for cell survival, growth, and other physiological processes [1]. Moreover, physical stimulation of cells is equally important and can be exerted in many different forms, such as an electric field, mechanical forces, dimensionality constraints, light or even a physical plasma [2–4]. One of the most prominent types of physical signals in organisms is the mechanical interaction between cells and their microenvironment. The ability of cells to sense forces and to further translate them into intracellular signals, as well as to exert forces back on the microenvironment, is called *mechanotransduction*. Cells can detect mechanical stimulation via several types of receptors, including the adhesion sites that are bound to the cytoskeleton [5]. Adherent cells adapt to the mechanical microenvironment through the dynamic rearrangement of their cytoskeleton and, in turn, exert the so-called *traction forces* back on the environment. Quantification of these forces can be achieved with Traction Force Microscopy (TFM) technique [6,7], which relies on registering the deformation of a cellular substrate and its subsequent recalculation into traction forces [8]. TFM became one of the most fundamental methods in cellular mechanobiology, explaining the role of substrate stiffness, cell contraction, migration, adhesiveness, and focal adhesion architecture in cellular mechanics [9,10] as well as investigating the role of mechanical signals in tissue dynamics [11,12] and cancer metastasis [13,14].

Another relevant physical factor that affects cells is temperature. Even in homeothermic organisms, the internal temperature varies owing to specific cell functions [15] and metabolism [16]. Furthermore, in vitro studies indicated that increased temperature can affect the myoblast proliferation [17]. Since thermal heterogeneity was observed in tumors [18] and at infection sites [19], temperature monitoring can be useful in diagnostics. Moreover, the stability of physiologically relevant temperature is crucial in live-cell microscopy as a primary factor for the research reliability and reproducibility. One of the most popular tools for temperature sensing are thermocouples, which are placed inside a Petri dish to measure the temperature of the cell medium. However, typical thermocouples are of millimeter size, which causes their relatively high thermal capacity, leading to a high inertia of their temporal response (~300 ms) [20]. Their dimensions also make them impractical for local microscopy measurements. These problems were partially solved by developing micro-thermocouples with significantly lower inertia and better measurement precision [21–23]. Recently, several groups have started the relatively new research field called *micro-* and *nanothermometry*, which focuses on local temperature measurements [24]. Among the many tools, one can mention the employment of quantum dots [25], organic fluorophores [26], biomolecules [27], synthetic polymers, [28,29] and diamonds. The latter have gained growing popularity in various branches of science, utilizing different types of color centers, including silicon-vacancy (SiV) [30] and nitrogen-vacancy (NV) centers [31] being the most popular.

The nitrogen-vacancy center in diamond is a defect in which one carbon atom is replaced by nitrogen and an adjacent lattice site is left empty, creating a vacancy. It may be electrically neutral ($NV^0$) or negatively charged ($NV^–$). $NV^–$ color centers have high photostability and exhibit strong red fluorescence after green-light excitation, making them useful as fluorescent markers in biological studies [32–35]. They are also paramagnetic, and hence can be controlled by the *Optically-Detected Magnetic Resonance* (ODMR) technique, which relies on the decrease of fluorescence intensity after application of a microwave (MW) field with an appropriate frequency around 2.87 GHz [36]. With the ODMR technique, one can use the $NV^–$ centers for sensing electric [37] or magnetic [38–40] fields. Owing to the linear temperature dependence of the ODMR frequency of approximately -74 kHz/K at room temperature [41] (Fig. 1A), they are also suitable as temperature sensors that can be used in biology [35,42–45].

Current studies of $NV^–$ nanodiamonds as temperature sensors in biology adopted various methodological approaches. Differences are primarily visible in microscopy techniques: some studies used confocal setups with point illumination, registering one diamond at time [42,45,46], while other used wide-field microscopes that register fluorescence from multiple diamonds [35,47]. In all described studies samples were illuminated with green laser light of intensity that varied from 4.7 $W/cm^2$ [45] for point illumination up to 3 $kW/cm^2$ for wide-field illumination [35]. The illumination intensity is crucial for live-cell experiments because intensive laser light may cause phototoxic damage, leading to further cell death. However, too low illumination decreases the contrast of the ODMR spectrum and thus the accuracy of the measured temperature. Another factor that influences ODMR quality is the type of fluorescence detector used. While the described confocal microscopes were equipped with avalanche photodiodes (APDs) [42,45,46], the wide-field setups were coupled with a single-photon detector [47] or a sCMOS camera [35], which also influenced the quality of the registered ODMR spectrum. Moreover, the resonance can be registered and processed in various ways as well. In some studies, the microwave sequence consisted of two MW frequencies from the descending fragment of the ODMR spectrum and another two frequencies from the ascending fragment, which were registered multiple times and then averaged. The resonance position was found at the intersection of two linear functions fitted to each two-point set. In these studies, data acquisition took milliseconds to hundreds of milliseconds, and the accuracy of the temperature measurement was in the range of hundreds of millikelvins for nanodiamond-based temperature registration [42,46]. In contrast, other studies measured the entire ODMR spectrum, which took from 12 s [35] up to 3 – 6 min [45]. Subsequently, the ODMR spectra were fitted with one or two Lorentzian curves, resulting in temperature accuracy ranging from millikelvins [35] to ~1 Kelvin [45]. The parameters of ODMR registration that influence acquisition time are crucial for live-cell imaging as well, since long exposure to microwaves and the antenna that heats itself can heat the sample up, and light illumination can induce phototoxic effects. Therefore, the resulting accuracy of temperature measurements in live-cell imaging generally originates from the trade-off between experimental conditions and cell viability.

Many experimental setups used in cell biophysics can measure only one physical factor at a time (such as substrate elasticity, mechanical deformation, application of electric current, etc.). Moreover, the authors of this work did not find any reports linking the mechanical response of living cells to local temperature. Therefore, our primary goal was to develop a wide-field microscope setup for parallel measurements of local relative temperature and cellular traction forces at once. This setup should be sufficiently simple to be applied in life science laboratories with a focus on short measurement times and sufficient precision for temperature measurement. With this in mind, we have built such a setup by incorporating microwave equipment into an existing commercial wide-field microscope equipped with a mercury short-arc fluorescence lamp, dry 40x/0.6 objective, and a CMOS camera without any extensive modifications of the microscope. An equally important aim was to decrease the potentially adverse impact of the ODMR experiment on living cells, possibly caused by intense light excitation, strong microwaves, and long exposure times. These conditions are typical in standard measurements in physics and material science, but are highly undesirable in experiments with living cells. One way to mitigate these effects is to keep the ODMR measurement as short as possible. The final objective was to achieve sufficient precision in the temperature measurements, which was approximately 1 K in our case. Such precision would provide physiologically significant information, making it a good starting point for further development of the proposed ODMR-TFM method.

In this work, we present the preparation of an ODMR-TFM experimental material, the optimization of ODMR measurements, and a demonstration of the exemplary combined ODMR-TFM measurements of cells that were heated or cooled down.

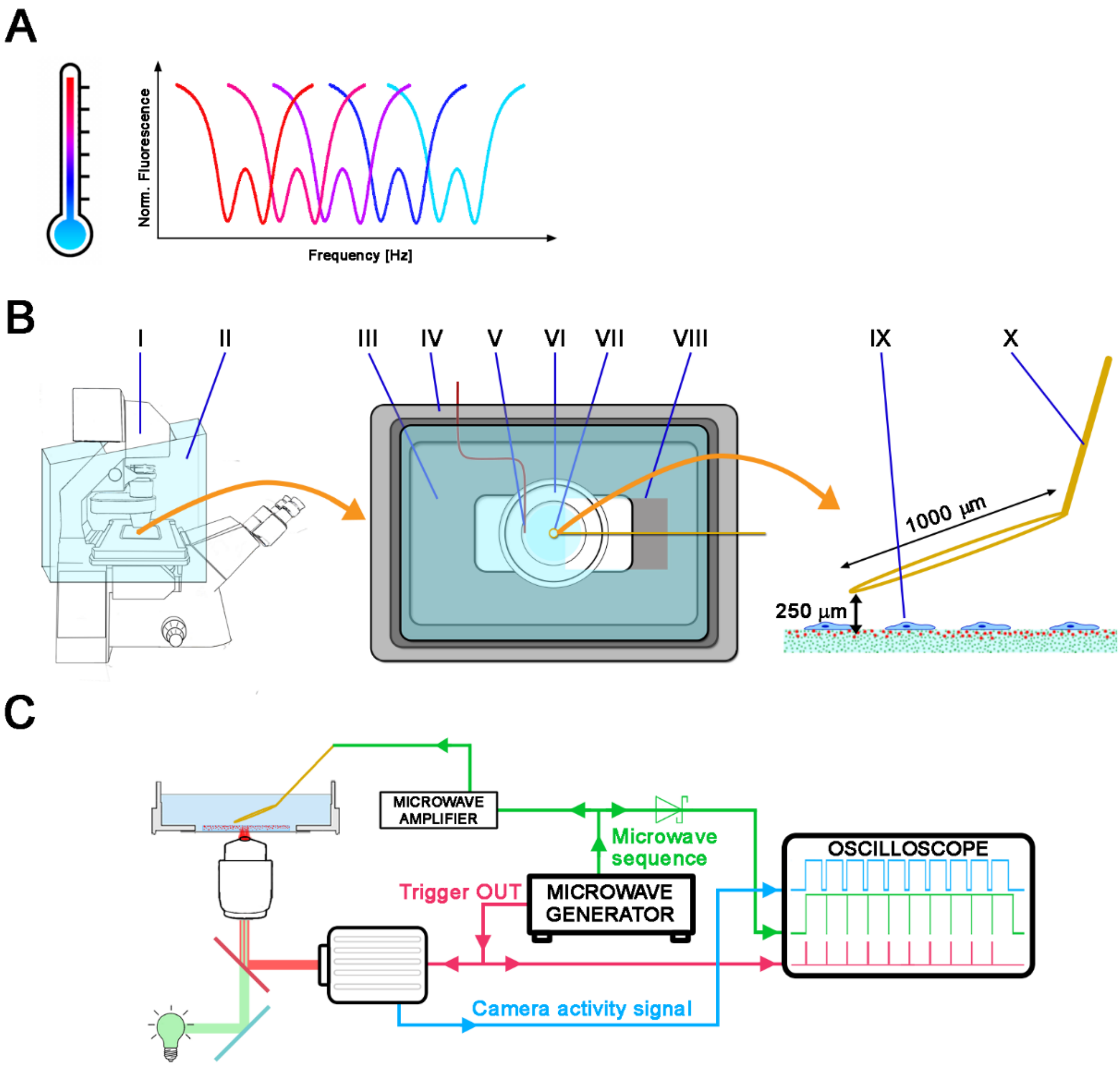

*Fig. 1. **A:** Idea of the ODMR temperature measurement: The ODMR resonance curves shift toward lower frequencies with increasing temperature. **B:** Scheme of the cell incubation equipment and the microscope setup: the left panel shows the fluorescence microscope (I) with the large incubation chamber (II). The middle panel presents the details of the incubator: III- heated mini-incubator on-stage (bottom), IV-incubator lid with window (top, not heated), V-thermocouple placed inside the inner well of the dish, VI-glass bottom dish with cells, VII-microwave antenna, VIII-window in the lid for the antenna. The right panel schematically depicts the side view of IX- observed cells and the X- antenna loop wire. **C:** Scheme of the microwave setup. The microwave generator serves as the microwave source as well as the camera trigger. All signals were observed on the oscilloscope.*

## Results

Our experiments were performed with a standard wide-field inverted microscope setup (Zeiss Axio Observer Z1) equipped with a short-arc mercury fluorescence lamp and a CMOS camera (Fig. 1 B-C). The sample was heated in two ways: by warm air in the large microscope heating chamber (Fig. 1B II) and by heating a small incubator for Petri dishes (Fig. 1B III), which was placed on the microscope stage. The temperature of the sample was controlled using built-in thermocouples and temperature controllers (Temp Module S and Heating Unit XL, Zeiss). The bottom part of the inner incubator was heated, while its lid (Fig. 1B IV) served as a partial passive heat barrier, containing a hole to introduce the microwave (MW) antenna above the sample. The actual temperature of the cell medium was measured using a thermocouple inserted into the dish on the left side of the well, touching its bottom (V). The MW antenna, in the form of a coaxial cable terminated with a wire loop of ~1 mm diameter made of thin copper wire (Fig. 1B VII), was placed on the micromanipulator, which was able to lower the antenna to ~250 μm above the substrate (Fig. 1B X). The MW generator served as the source of the microwaves and as a camera trigger (Fig. 1C). Three signals (MW sequence, camera activity signal, and camera trigger) were observed on oscilloscope to verify the correctness of the ODMR measurements (Fig. 1C).

Our work on the combined ODMR-TFM setup consisted of three stages: optimization of the ODMR-TFM substrate (Fig. 2), optimization of the ODMR measurement conditions (Fig. 3), and realization of the exemplary ODMR-TFM experiments (Fig. 4).

*Development and characterization of an ODMR-TFM substrate*

For the first task, a polyacrylamide (PA) hydrogel of defined elasticity containing both diamonds and fluorescent markers was prepared. To register the ODMR signal under low illumination, short experiment time, and low-aperture objective, we decided to use microdiamonds (µDs) of ~1 µm diameter that contain a large number of $NV^-$ centers (~$10^5$ centers in each microdiamond) emitting light in the red channel. However, such large markers are rather not suitable for TFM experiments, and therefore we also applied green fluorescent polystyrene (PS) beads 200 nm in diameter to register cellular tractions in the green channel of the microscope.

To keep the µDs properly dispersed within the substrate, it was vital to test their surface termination in various solutions. Primarily, we tested oxygenated and hydrogenated nonfluorescent µDs in four different solvents (Fig. 2A, Table 1): deionized (DI) water (suitable for PA polymerization mixture), DMSO (potentially dissolvable in PA polymerization mixture), and two protein solutions: bovine serum albumin (BSA) 2.5% w/w in water and fetal bovine serum (FBS). The latter two solvents were chosen based on previous studies, indicating that the creation of a protein corona can prevent the aggregation of nanodiamonds [48]. The results of dynamic light scattering (DLS) measurements are presented in Fig. 2A and Tab. 1. and show that most combinations of surface termination and solvents worked properly. The only exceptions were oxygenated µDs in DMSO and hydrogenated diamonds in FBS, which had a significantly higher hydrodynamic diameter than other samples. However, it should be noted that the increased hydrodynamic diameters of µDs in protein solutions could also be caused by the creation of the protein corona; thus, they do not necessarily signify the actual aggregation of µDs. Since the lowest hydrodynamic diameter was observed for the oxygenated µD suspension in water rather than in protein solutions, we decided to use the water suspension in further experiments. $NV^-$ µDs were oxygenated and suspended in DI water at a concentration of 0.5 mg/ml. The DLS measurement showed correctly dispersed µDs with a fraction of large aggregates (> 6 µm in diameter) that constituted less than 10% of the total measured DLS intensity. This suspension of oxygenated µDs in DI water was also used to prepare elastic substrates.

Substrates suitable for traction force microscopy must have a properly characterized value of Young modulus. As shown in Fig. 2B, we compared the three applied substrate formulations that should lead to the a targeted elasticity of ~12 kPa. The substrate without fluorescent particles had an elasticity of 11.7±0.5 kPa, similar to those found in the literature [49]. The substrate with added µDs was slightly softer with a Young modulus of 10.8±0.4 kPa. The addition of fluorescence markers softened the substrate to 9.6±1.6 kPa, which was the final value considered in our TFM calculations. . However, even though the ODMR-TFM substrate with µDs and green fluorescent polystyrene (PS) beads is slightly softer than the substrate without added particles, it still has acceptable elasticity and is suitable for mechanobiology studies. Another aspect of the substrate is the surface morphology. AFM experiments showed that the substrates with µDs exhibited shallow nanopits up to 140 nm in depth (Fig. 2C-D). The 3D morphology of the ODMR-TFM substrate is presented in Fig. 2E, which shows that fluorescent microdiamonds are placed in the top layer of the substrate; however, each of them might be at a slightly different depth. In contrast, fluorescent PS beads are rather gathered at the bottom part of the substrate; however, their concentration at the top layer was sufficient for the TFM experiments. Such a diverse placement of microdiamonds and PS beads might be caused by the upside-down polymerization during which the heavy microdiamonds sedimented into the bottom and low-density PS beads floated to the top of the substrate. Reversing the substrate back to the upright position after polymerization resulted in the placement of microdiamonds in the top layer and PS beads in the bottom part of the substrate. We expect that the appearance of pits on the substrate was caused by the presence of diamonds beneath it. This hypothesis can be verified by a possible combination of AFM and super-resolution microscopy, which is beyond the scope of this study. In summary, we showed that we were able to produce a substrate with properly dispersed microdiamonds, which was slightly softer than the reference substrate without any particles, but still well suited for the planned TFM and ODMR measurements. The nanopits existing on the substrate surface were rather shallow; however, they can be considered in further studies.

*Table 1. DLS characterization of the hydrodynamic diameters of nonfluorescent (non-$NV^-$) and fluorescent ($NV^-$) diamonds. Quantitative characterization presents the average hydrodynamic diameter ($Z_{Ave}$), standard error of the mean (SEM), and the 10th, 50th and 90th quantiles ($Di_{10}$, $Di_{50}$, $Di_{90}$).*

| $NV^-$ presence | Surface termination | Solvent | $Z_{Ave}$ [nm] | SEM [nm] | $Di_{10}$ [nm] | $Di_{50}$ [nm] | $Di_{90}$ [nm] |
|---|---|---|---|---|---|---|---|
| Non – $NV^-$ | -O | DI water | 734 | 88 | 546 | 754 | 1060 |
| | | DMSO | 1145 | 279 | 813 | 1220 | 1860 |
| | | BSA 2.5% | 741 | 108 | 540 | 767 | 1100 |
| | | FBS | 887 | 168 | 597 | 945 | 1540 |
| | -H | DI water | 818 | 187 | 578 | 864 | 1310 |
| | | DMSO | 798 | 108 | 568 | 817 | 1200 |
| | | BSA 2.5% | 851 | 192 | 548 | 907 | 1570 |
| | | FBS | 1267 | 385 | 942 | 1430 | 2230 |
| $NV^-$ | -O | DI water | 734 | 202 | 484 | 713 | 1260 |

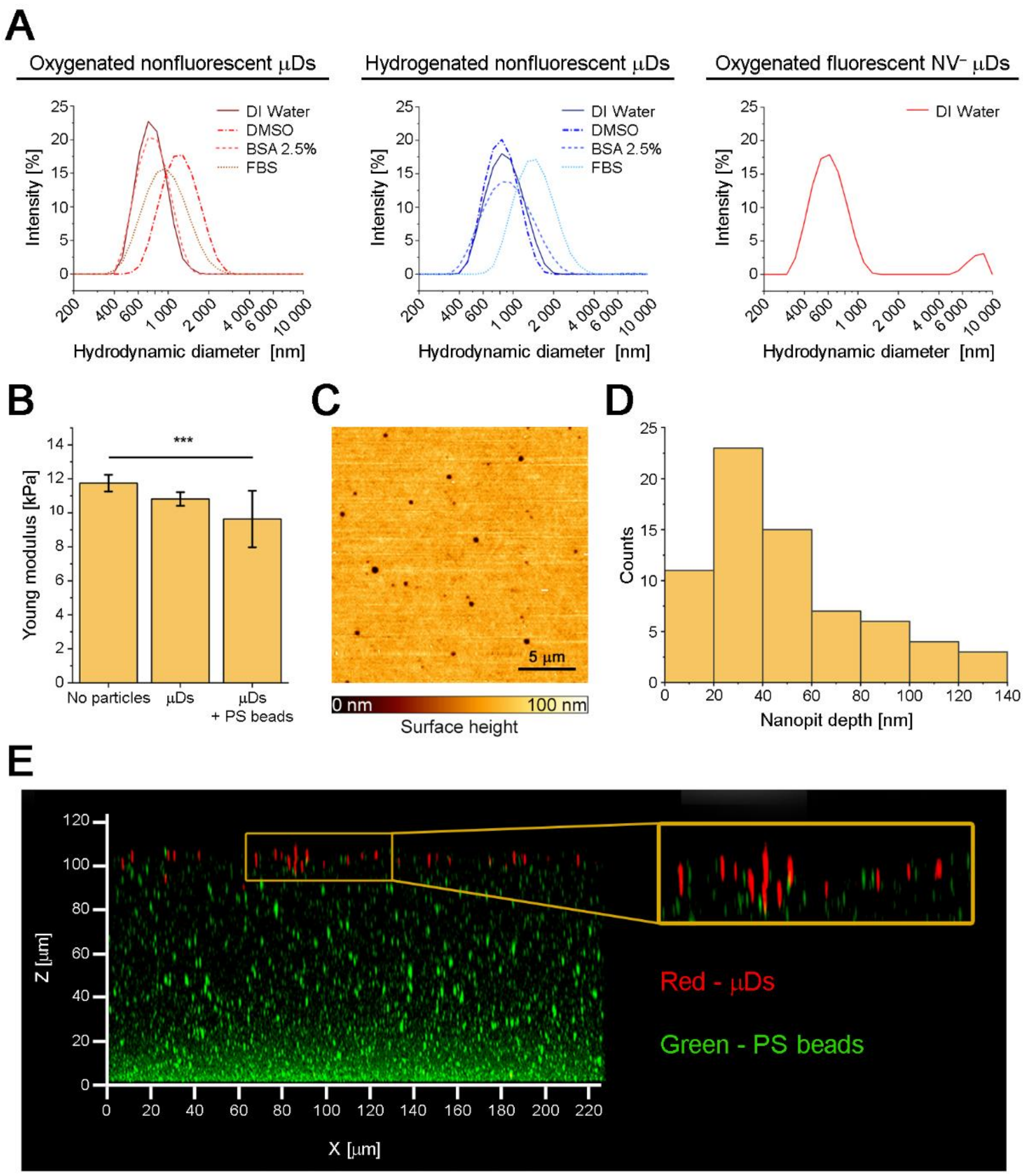

*Fig. 2.* ***A:*** *Hydrodynamic diameters of oxygenated and hydrogenated nonfluorescent and fluorescent microdiamonds suspended in different solvents.* ***B:*** *Young's moduli of different variants of elastic substrates.* ***C:*** *Surface morphology of the ODMR–TFM substrate.* ***D:*** *Histogram of nanopit depths.* ***E:*** *Maximum projection of the ODMR–TFM substrate in the X-Z plane from a confocal scan. The yellow box presents the magnified fragment of the substrate surface, showing that the microdiamonds are placed at a slightly different depth.*

*Optimization of ODMR measurement conditions*

The next step in developing the ODMR-TFM setup involved optimizing the experimental conditions of the ODMR. The observation presented in Fig. 3A shows that the ODMR spectrum is split into two components, which are caused by diamond-lattice strain [36]. The current literature presents several ways of determining the position of ODMR, and thus, the readout of the relative temperature, as mentioned in the Introduction. Here, we found (Fig. 3A) that, rather than fitting two Lorentzian components, we can achieve more accurate resonance localization while fitting a single Lorentzian shape to the central dip between the strain-split ODMR components. Such a procedure enables not only a sufficiently accurate identification of the resonance frequency but also significantly shortens the required microwave sequence, reducing sample heating.

Assuming that live-cell imaging will be performed in the range of 20-45°C, we decided to probe the ODMR spectrum in the range of 2864 and 2874 MHz with a 0.25 MHz step with three additional points outside the resonance at the beginning and at the end of the spectrum. In this work, we call this MW sequence an "irregular sequence". We have compared it with a reference microwave sequence (called the “Regular Sweep” here) that was performed between 2845 and 2895 MHz with the same 0.25 MHz step. All these measures served to decrease the duration of the ODMR measurement as much as possible while maintaining the ~1K accuracy.

Fig. 3B presents the procedure for determining the average relative temperature in the field of view developed in this study. The first step consists of obtaining the ODMR spectra for each visible particle. This is followed by the selection of particles (NV1, NV2, NV3, etc.) that produce ODMR signals of sufficient quality to allow a reliable determination of the resonance frequency, that is, (a) at least 1% of the height of the dip $H$, (b) the height uncertainty $\delta H$ not larger than the height of the central dip itself $\delta H/H < 1$, and (c) the $R^2$ factor of the fitted Lorentzian should be ≥0.4. These rather low requirement criteria were chosen to include even the weakest ODMR signal in our analysis. The selected signals were averaged, and a Lorentzian was fitted to the averaged data to determine the resonance position. Microdiamonds with insufficient quality ODMR signals were not considered in further analysis. The low quality of the ODMR could have been due to the small size of the µD or the fluorescence of a defocused diamond. This procedure allowed the identification of the relative temperature in consecutive time steps.

To determine the best set of conditions for ODMR measurements, we examined the accuracy of the temperature readings and number of diamonds included used in the analysis based on three parameters: a) the thickness of the substrate, b) the intensity of the optical excitation, and c) the type of microwave sequence. For the thickness of the substrate, we tested thin substrates (between 20 and 40 µm) that were too thin for standard cell mechanobiology measurements, medium-thick substrates (between 50 and 70 µm) that are just above the reliability threshold in the context of mechanobiology studies, and thick substrates (between 100 and 120 µm) that are of comfortable thickness for cell mechanobiology studies. The excitation intensities were chosen from the maximum available intensity of the fluorescence lamp (25 W/cm$^2$), which was gradually reduced in 5 W/cm$^2$ steps. For the microwave sequences, we compared the Irregular Sequence that probed only the central ODMR dip with the Regular Sweep. We also determined the influence of averaging the Irregular Sequence: 1, 2, 3, and 4 times, which took 1.2 s, 2.4 s, 3.5 s, and 4.7 s, respectively. This comparison was used to find the shortest possible sequence that provides sufficient temperature determination accuracy and includes a reasonable number of diamonds, which also increases the reliability of the temperature measurement (temperature is averaged from more places in the field of view).

Our results show that the accuracy of the temperature measurement increases with substrate thickness. This result may come from the fact that in our assay we used a glass bottom dish of #0 thickness, while the objective was calibrated for a standard #1.5 coverslip. This might cause defocusing of the imaged fluorescence signal for thin substrates but should be negligible for thicker ones. A satisfactory temperature measurement precision of 0.76±0.16 K (which also achieved the target accuracy of <1 Kelvin) was obtained for illumination of 15 W/cm$^2$ and the Irregular Sequence averaged twice. It has also delivered the 12.6±2.07 analyzed diamonds that met the quality criteria of the ODMR spectrum. This optimization allowed us to decrease the time of the ODMR experiment to 2.4 seconds and 15 W/cm$^2$ excitation, while using comfortably thick PA substrates and maintaining the proper accuracy of the temperature measurements.

The temperature response of the microdiamonds was also verified (Supplementary Figure S1 and Supplementary Table S1). The temperature calibration shows slopes between -65.6 kHz/K and -71.2 kHz/K, which are similar to those found in the literature. This confirms that our ODMR-based temperature setup exhibits appropriate linear characteristics. However, for the relative temperature change calculation, we used the slope value of -74.2 kHz/K from Acosta et al. [41] as measured under more precisely controlled conditions than those in our microscopy setup.

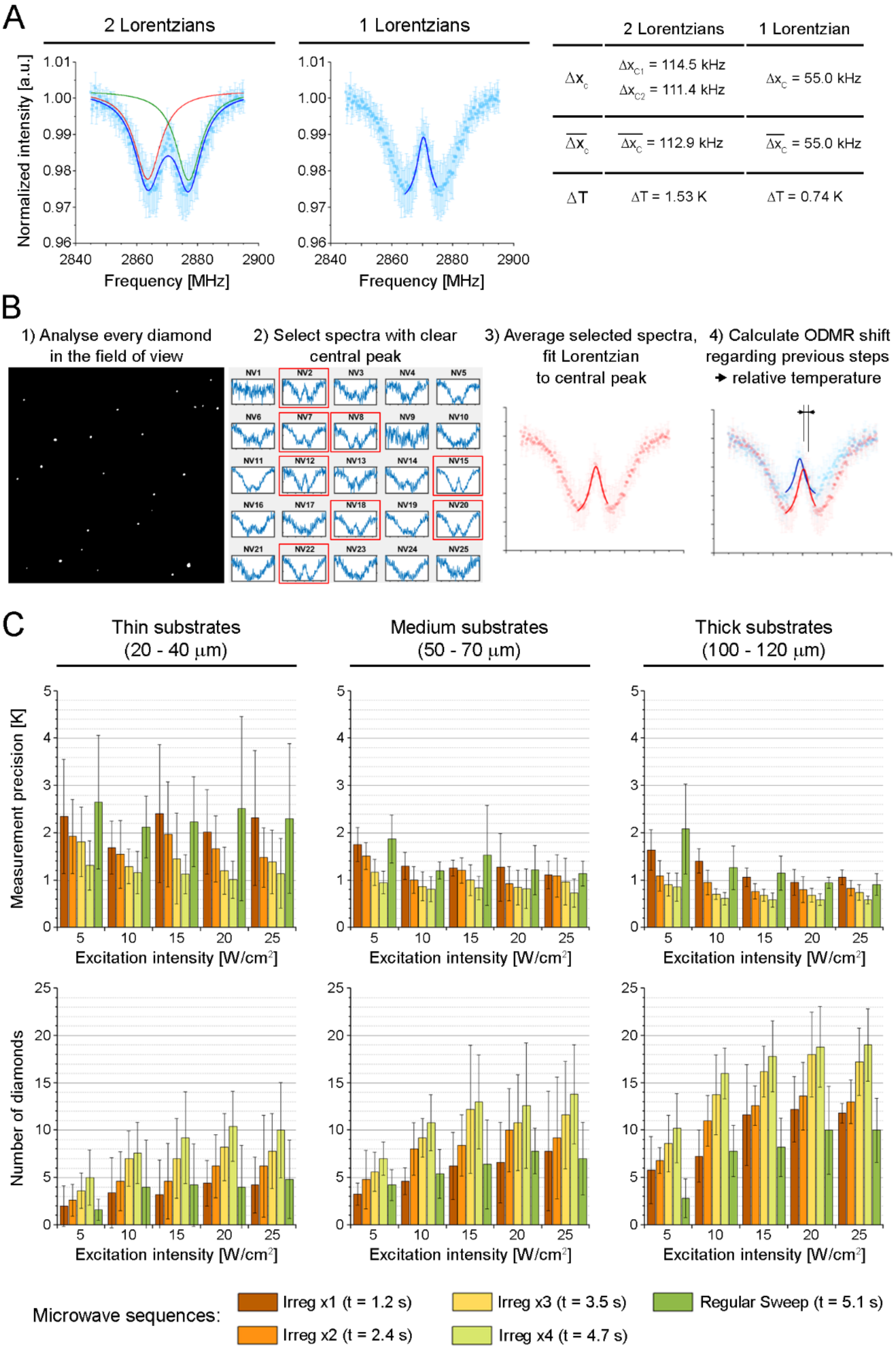

*Fig. 3.**A:** Accuracy of the determination of the center of ODMR spectrum fitted by two Lorentzian functions and one Lorentzian function. The uncertainty of temperature was calculated assuming slope -74.2 kHz/K, based on Acosta et al. [41]. **B:** Idea of calculating the relative temperature that is averaged in a field of view. **C:** Optimization of the ODMR experiment. Top row: influence of experimental parameters on the accuracy of temperature measurement. Bottom row: influence of experimental parameters on the number of diamonds included in the analysis.*

Finally, two exemplary experiments for cooling and heating were conducted. Both 4-hour experiments started with maintaining a stable temperature for 1 hour, which was followed by a gradual change of temperature by ~ 1 K every 10 minutes. The final temperature was maintained until the end of the experiment. The correct final temperature of the whole sample during heating and cooling was determined by the thermocouple readout (bulk signal from the inside of the dish). The cooling experiment started at 37.0°C, while the heating experiment started at 28.4°C. The upper panel of Fig. 4A presents the temperature measured by μDs (red squares) and the thermocouple (blue circles), while the lower panel shows the constrained tractions exerted by the cells (green bars) during the experiments. Since the values of the constrained tractions from the map did not meet the normality criteria, they are described here by mean , median, and quartiles at each time point. The time points highlighted in purple in the plots in Fig. 4A are presented in Fig. 4B in more detail.. The thermocouple temperature measurement for sample cooling shows a rather smooth decrease, while the μD measurement shows temperature drops at 70, 120, and 160 min, which were then followed by temperature stabilization. We regard the latter measurement as more reliable because the temperature drops coincide in time with the addition of the next portions of ice packs into the large incubation chamber to cool down the microscope environment. Furthermore, the temperature measured by the μDs had consistently lower error bars than the temperature measured by the thermocouple. Similarly, the heating experiment yielded more precise temperature measurements using μDs (smaller error bars). They also responded to the heating of the sample, showing the dynamics of temperature over the course of the experiment. Each experiment consisted of maintaining a constant temperature for the first and last hour. Even if the thermocouple readouts were stable in these regions, the microdiamond readout exhibited temperature fluctuations. Owing to the faster registration of the ODMR compared with the thermal inertia of the thermocouple, we can assume that the microdiamond registration is more reliable than the bulk thermocouple signal. In the lower panel of Fig. 4A, one can see that the distribution of constrained cellular tractions in heated cells is visibly wider. This comes from the larger area of the observed cell, which exerts larger forces near the cell border and includes more pixels of low tractions closer to the cell center. While Fig. 4A presents a quantitative description of constrained tractions, Fig. 4B shows unconstrained tractions maps to better assess the quality of the TFM experiments. The maps of unconstrained tractions for cooling and heating show sufficient quality and low background noise of registered tractions, and demonstrate that parallel measurements of relative temperature and cellular tractions are possible in our setup. The ODMR spectra at each selected time point explicitly show a spectral shift to the higher frequencies for cooling and lower frequencies for heating the cells.

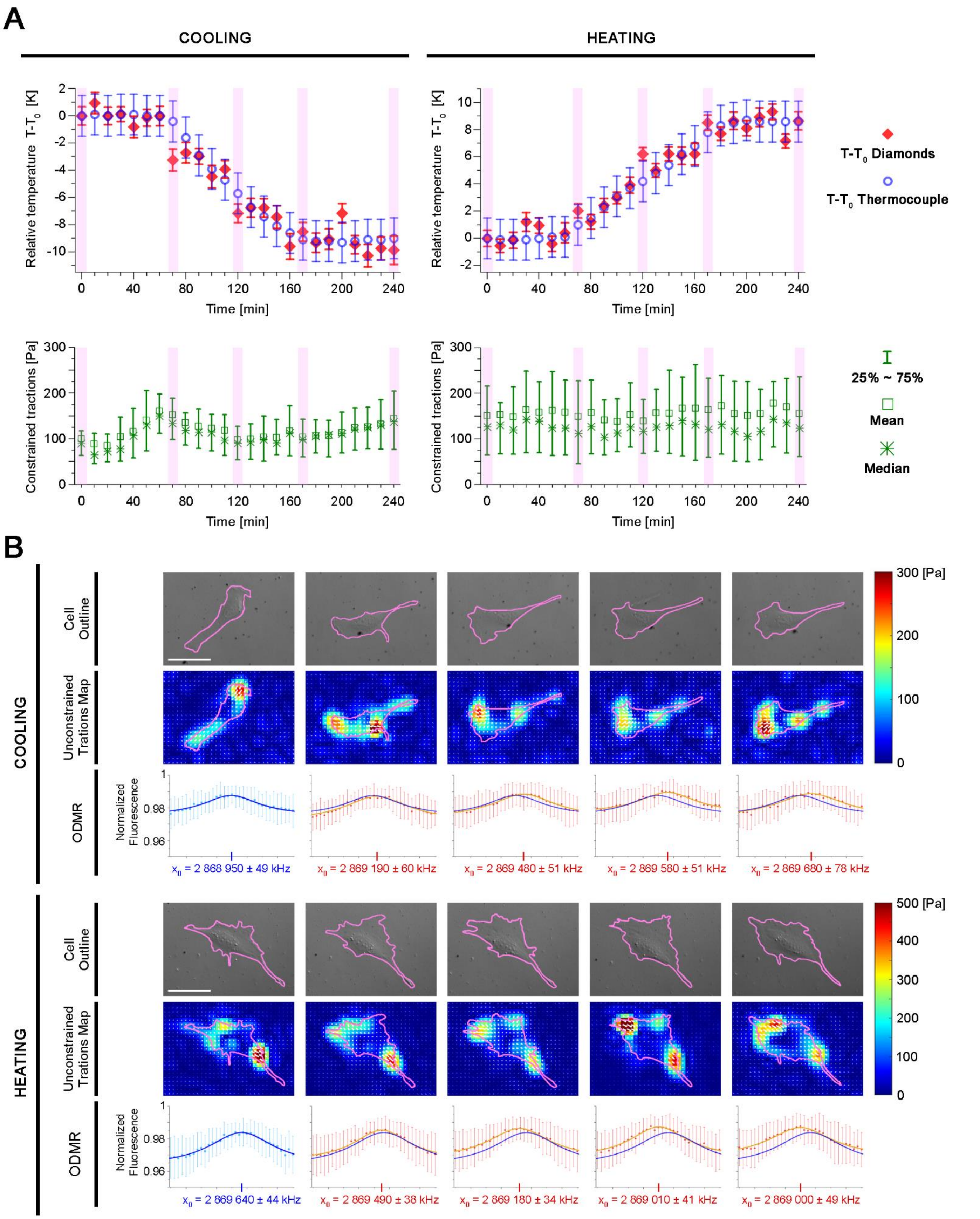

*Fig. 4. ODMR-TFM proof-of-principle experiments for cell cooling and heating.* ***A:*** *Plots of relative temperature measured with the thermocouple and microdiamonds (upper panel) and constrained tractions exerted by the observed cells (lower panel). The timepoints highlighted in purple in the plots are presented in more detail in panel B.* ***B:*** *Cell outlines, maps of unconstrained tractions, and ODMR shift for cooling and heating. Scalebar = 50 µm.*

## Discussion

In this study, we developed a microscope imaging method for the hybrid sensing of cellular tractions (TFM) and local relative temperature (ODMR). In addition to the proof-of-principle experiment, we described the optimization of ODMR measurements. The constructed setup is relatively simple and is based on a standard wide-field fluorescence microscope without extensive modifications, while the experimental conditions were optimized to decrease the harmful effects of light and microwaves on living cells.

The current research began with the preparation of an ODMR-TFM substrate that contained fluorescent beads for TFM and microdiamonds for ODMR. The substrate was tested for its elastic properties, which are crucial factors in TFM experiments [6,7], and topography, which might affect cellular behavior. The developed custom ODMR-TFM substrate slightly differs in elasticity from the reference one, however, it does not change its applicability in TFM experiments. It only requires a modification of the Young modulus value, which is further used to calculate cellular tractions. In further studies, it is also important to verify whether the Young modulus of ODMR-TFM substrates of different elasticities is not changed. The substrate was also sufficiently thick (>100 µm) to allow mechanobiological studies. Substrate surface topography imaging revealed the presence of shallow nanopits (up to 140 nm). Current literature shows that nanopits can modify cell spread, adhesion, differentiation, or osteogenic induction on substrates [50–52]. However, in the described cases, the nanopits were deeper, regularly and densely packed together. In our study, the nanopits are shallower and sparsely located at random places on the substrate, so we do not expect them to have a significant impact, although one might consider them in further studies if cell behavior is somehow altered.

In our study, microdiamonds were not introduced directly into the cell, as in Kucsko et al., [42] or by culturing with nanodiamond suspension [35], but were placed in the hydrogel, just below the cells. This approach does not provide any additional interaction between cells and diamonds, thus reducing the effect of external factors acting on cells. Even if diamond particles exhibit high biocompatibility [53] and ref. therein, the lack of additional direct interactions makes the experiment less prone to perturbations.

Another objective of this research was to optimize the ODMR experimental conditions in the context of live-cell imaging. The microthermometry of living cells using the ODMR method is a relatively new topic. Therefore, only a few studies have presented the relevant experimental approaches. Similarly, there are no standard live-cell ODMR protocols, and experimental conditions have rarely been optimized up to now. In this study, we optimized the microwave sequence for ODMR registration, data processing, illumination intensity, and substrate thickness. This was the result of a relatively short experiment (2.4 s) that used low illumination (15 W/cm2), providing satisfactory accuracy in temperature measurement (0.78±0.16 K). The temperature accuracy is similar to that reported by Yukawa et al. [45] however, in our study, it was achieved with a simpler (nonconfocal) setup, significantly shorter experiment, but slightly higher illumination (4.7 W/cm$^2$ in [45]). However, the temperature accuracy achieved in our work is worse than that reported in another wide-field setup with sCMOS camera that was developed by Simpson et al. [35]. They have achieved a temperature determination accuracy that ranged in millikelvins; however, they also applied 200 times more intensive illumination (3 kW/cm2), and the experiment time was four times longer than that presented in our work. Other studies [42,46] have also reported better temperature accuracy; however, they utilized more sophisticated confocal microscopes that probed only a small fraction of the field of view. In summary, the ~ 1 K temperature accuracy achieved in this work is not the highest that can be found in the literature; however, it still provides relevant biological information about the thermal conditions of cells, including information from several diamonds in the field of view, and was achieved with low illumination, short experiment time, and on the relatively simple wide-field microscope setup that does not require major modifications.

It is clear that the final quality of the temperature measurement is derived from the interplay of several experimental conditions and the microscope equipment. To the authors' knowledge, this is the first systematic study to optimize the experimental conditions of live-cell ODMR. However, the parameters optimized here can still be examined further to increase the accuracy even more, while decreasing the heating from microwaves and phototoxic effects due to illumination. Therefore, future studies should aim to determine the extent to which light or microwave perturbations influence cell functions. In addition, the biological effects of the ODMR experiment on living cells, such as the creation of free radicals or the induction of the apoptotic signaling pathway, were not measured here, whereas the optimization procedure aimed to keep the observed cells alive during the several-hour experiment. Therefore, explicit registration of the negative biological effects of an optimized ODMR experiment remains an open task. Addressing this issue can further enhance the applicability of ODMR to live-cell imaging. Another improvement can be achieved by modifying the MW antenna design and delivery. Now, the antenna was positioned from the top of the sample, which imposes an open type of sample container. By positioning the antenna in a manner that maintains the enclosure of the sample, the robustness of the experiment can be increased. However, this adjustment may introduce additional complexity into the experimental setup.

To the authors' knowledge, this work is also the first to apply local temperature measurements in the field of mechanobiology. Proof-of-principle ODMR–TFM experiments showed that it is possible to observe the local temperature in the field of view and cellular traction forces using a relatively simple microscopy setup. This method can be used in studies of hyper- and hypothermic conditions or even rapid temperature changes in the mechanical response of cells and tissues, as $NV^-$ diamonds have low thermal inertia and provide a precise temperature readout from the observed field of view. Therefore, coupling more than one method allows deeper insight into cell biophysics by the parallel observation of one more physical experimental condition simultaneously.

## Materials and methods

*Elastic substrate preparation*

Elastic substrates were prepared in two forms: first, in a glass-bottom dish (CellVis, #0 thickness) for microscopic observations and on standard glass slides for AFM measurements. For both types of substrates, glass was silanized using a mixture of 3-(trimethoxysilyl) propyl methacrylate (Sigma-Aldrich) with glacial acetic acid (Chempur) and 96% ethyl alcohol in 1:1:14 proportions for 20 min. After this process, the glass was rinsed three times with 96% ethyl alcohol to wash our any residues of chemicals [49]. The polymerization mixture of the blank 12 kPa substrate was prepared by mixing 7.5% acrylamide and 0.16% bis-acrylamide in DI water; mixture of the 12 kPa substrate with diamonds was prepared similarly to the blank substrate but with the addition of the pre-sonicated and vortexed suspension of NV– diamond at 25 µg/ml concentration. The polymerization mixture of the ODMR–TFM substrate (substrate for parallel measurement of temperature and cellular traction) was prepared similarly to the substrate with NV– diamonds, but with the addition of 2% green fluorescent, 200-nm diameter carboxylate-modified polystyrene (PS) beads (excitation: 505 nm, emission: 515 nm, Thermo). Each type of polymerization mixture was mixed with polymerization agents: 0.05% ammonium persulfate and 0.05% tetramethylethylenediamine (Bio-Rad). All substrates were prepared by upside-down polymerization, and the appropriate amount of polymerization mixture (18 µl for thin substrates, 22 µl for medium, and 26 µl for thick substrates) was placed on the glass coverslip, which was then touched from the top by the inner surface of a glass bottom dish or a glass slide to allow upside-down polymerization. This procedure allowed for the sedimentation of microdiamonds. After 1h of polymerization, the upside-down glass bottom dish was turned orderly and filled with PBS, allowing detachment of the top coverslip (which was initially at the bottom). After polymerization, the substrates were left for 24 h to dissolve any residual acrylamide. Subsequently, the substrates were washed three times with PBS.

*Diamond surface functionalization and particle size characterization.*

Tests of different surface terminations in various solvents were conducted on nonfluorescent microdiamonds (MSY 0.75-1.25, Pureon AG). They have a size (D50 = 1 µm) and surface properties similar to those of $NV^-$ microdiamonds but are less expensive. The final suspension used in further experiments was prepared from $NV^-$ microdiamond powder with ~3.5 ppm color centers concentration (MDNV1umHi30mg, Adámas Nano). Hydrogenated and oxygenated surface terminations were prepared by treating diamond powders with hydrogen and oxygen plasma using a plasma cleaner system (Diener ZEPTO). Briefly, the weighed diamond powder was placed in a glass dish in the chamber of the plasma cleaner system. After the air was pumped out, the chamber was filled with hydrogen or oxygen to hydrogenate or oxidate the surface, respectively. At a pressure of 0.3 mbar, plasma RF with 13.56 MHz frequency and 300 W power was employed. Diamond powder was treated with plasma for 10 min.

Next, four separate nonfluorescent microdiamond suspensions were prepared at a concentration of 0.5 mg/ml using the following solvents: (I) deionized water (prefiltered through a 0.22 µm PS filter), (II) DMSO, (III) BSA 2.5% w/w in DI water (freshly prepared from lyophilized BSA and prefiltered DI water), and (IV) FBS solution (FBS qualified, Gibco). Suspensions of $NV^-$ oxygenated microdiamonds were also prepared in prefiltered DI water at a concentration of 0.5 mg/ml. Before the DLS measurements, the samples were sonicated for 5 min in an ultrasonic bath and then vortexed twice for 15 s at 3000 rpm.

The hydrodynamic diameters were determined using a Zetasizer Nano ZS particle analyzer (Malvern Panalytical, UK) equipped with a 632.8 nm laser and a narrowband filter (ZEN9062). For measurement, each suspension was placed in a disposable polystyrene cuvette (outer dimensions of 12 × 12 × 45 mm) and measured at 25°C using a backscatter configuration (scattered light collected at a 173° angle). Each measurement of the particle size distribution consisted of three measurements. The particles were analyzed in the context of their mean hydrodynamic diameter (Z-Average), standard error of the mean (SEM), and the 10th, 50th, and 90th percentiles ($Di_{10}$, $Di_{50}$, $Di_{90}$) derived from the intensity-weighted particle size distributions. SEM was calculated from the standard deviation derived from the polydispersity index [54].

*Characterization of ODMR–TFM substrates.*

Elasticity measurements were performed using a NanoWizard 3 NanoScience AFM (JPK Instruments) in force mapping mode. The experiments were repeated three times on samples prepared independently on different days, and each experimental day consisted of examining all types of 12 kPa substrates. For each substrate, a force mapping was performed in five different areas of the sample. All experiments were performed in a drop of Hanks' Balanced Salt Solution (HBSS, Cat. No. 55037C, Sigma-Aldrich, St. Louis, USA). A spatial map of the force vs. distance (FD) curves was measured on a 16 × 16 point grid and a square surface of 30 µm × 30 µm. The position of the scanned area was controlled by an inverted optical microscope (IX71, Olympus). Force-distance curves were measured at a speed of 2 µm/s with a maximum applied force of 1.2 nN. To evaluate the elastic modulus of the substrates, a non-covered spherical polystyrene probe with a radius of 2.5 µm (Novascan) mounted on a triangular cantilever with a spring constant of 0.03 N/m was used. Before each measurement, the cantilever spring constant was calibrated using a dedicated software (SPM Software, JPK Instruments). The global elastic modulus of the substrates was calculated using the Hertz-Sneddon model.

High-resolution topographic images were obtained using force-distance (FD)-based imaging mode (QI; JPK Instruments). In this method, a single FD curve measurement is performed at every pixel of the image and then translated from the selected trigger force into images of the substrate topography. The loading force ranged from 0,5 to 0,7 nN and was adjusted to obtain a clear contrast of the topography of the substrate. The topography images obtained were analyzed using the JPK Data Processing Software.

The nanopit analysis started with a manual outline of the nanopit area and finding its deepest point. The nanopit depth was compared to the closest surrounding of the nanopit, that is, the envelope with a width of three pixels. Confocal imaging was performed using a Zeiss LSM 710 confocal microscope with a Zeiss α-Plan Apochromat 63x/1.46 oil objective.

*ODMR registration setup.*

The microwaves were generated by a generator (Rohde & Schwarz SMBV100A) connected to an amplifier (Mini-Circuits ZRL-3500+) linked to the MW antenna in the form of a coaxial cable terminated with a wire loop of ~1 mm diameter made of 0.2 mm copper wire. The registration of wide-field ODMR consisted of a time-lapse sequence of fluorescent images acquired for each next MW frequency with a CMOS camera (ORCA-Flash 4.0 V2). The camera was triggered by the signal from the MW generator, and therefore, each next frequency of the microwave sequence triggered the acquisition of each next image. To increase the camera registration speed, the ODMR spectrum was registered with a field of view limited to 1024 × 1024 pixels. The regular ODMR sweep (Regular Sweep) consisted of 201 discrete frequencies with a 0.25 MHz step between 2845 and 2895 MHz. The Irregular Sequence (in short, “Irreg”) consisted of the of 0.25 MHz-steps between 2864 and 2874 MHz with three off-resonance frequencies at the beginning and the end of the sequence. The MW frequency was switched every 25 ms. However, owing to the camera dead time, fluorescence was registered only for 18 ms. The MW antenna was positioned ~250 µm above the substrate, using the InjectMan® NI 2 (Eppendorf) micromanipulator. $NV^-$ diamond fluorescence was collected using a custom fluorescence cube built with the Band Pass Filter 470/40 nm (excitation), DMLP 567 nm dichroic mirror, and FELH 600 nm filter (emission).

*Optimization of ODMR experiments*

The optical power density of the fluorescence lamp was measured as follows: the LD Plan-Neofluar 40x/0.6 objective (Zeiss) was focused on the top layer of the ODMR–TFM substrate to obtain a sharp image of green fluorescent beads; the field diaphragm on the fluorescence path was partially closed to make its borders visible on the image, allowing measurement of the area of illuminated field of view (this area was then used to calculate the optical power density); the ODMR–TFM substrate was removed from the microscope stage and replaced with a power meter (PM100D power meter with S120 sensor, Thorlabs). The objective was refocused to register the maximum optical power of the focused light. The optical power was measured between 0 and 100% of the intensity of the fluorescence lamp (HXP120, Zeiss). The optical power registered by the power meter was then divided by the field of view area measured at the initial stage.

The thickness of the substrate was measured by refocusing between the top and bottom surfaces of the substrate using a dry objective (Zeiss, LD Plan-Neofluar 40x/0.6). Because of the differences between the refractive index of the PA substrate and air ($n_{air}$), the measured distance ($\Delta h_{Measured}$) was corrected using the formula below, as in the previous literature [55]. The refractive index of the polyacrylamide gel was assumed to be the same as that of water ($n_{water}$), and *NA* was the numerical aperture of the objective.

$$\Delta h_{Corr} = \Delta h_{Measured} * \sqrt{\frac{{n_{water}}^2 - NA^2}{{n_{air}}^2 - NA^2}}$$

The ODMR spectra for the three substrates of different thicknesses were collected at room temperature. They were used to examine five locations of similar thicknesses, each for three different thickness ranges: thin (20-40μm), medium (50-70μm), and thick (100-120μm).

*Temperature calibration of ODMR–TFM substrates*

Temperature calibration of the ODMR–TFM substrate was performed in the range of 17°C – 39°C. The ODMR–TFM substrate was placed on the same incubation equipment, as shown in Fig. 1B. A type K thermocouple of ±1.5°C accuracy was placed inside the glass bottom dish, on the left side of its inner well, touching the bottom coverslip. During the experiment, a glass bottom dish with an ODMR–TFM substrate was filled with the cell culture medium without phenol red to preserve the standard experimental conditions. Then, the temperature inside the dish was increased by 1°C, which was confirmed by a stable thermocouple readout for 5 min. This allowed us to assume that the glass-bottom dish with the substrate and cell culture medium had the same temperature. ODMR spectra were collected from three different fields of view. For each time point, the reference full sweep and chosen Irregular Sequence averaged twice (see: ODMR registration setup) were measured. To avoid the potential local heating of diamonds, these two measurements were performed at 1-minute interval.

*Cell culture*

MEF 3T3 and HEK293 cells were cultured in plastic T-25 bottles (NEST Biotechnology) in a standard cell culture incubator. MEF 3T3 were maintained in DMEM Low Glucose medium (Bio West) supplemented with 10% of FBS (Gibco) and 1% PS (BioWest).

*ODMR–TFM experiment*

Prior to the ODMR–TFM experiments, the surface of the 12 kPa PA substrate was functionalized with type-I collagen. The polyacrylamide substrate was treated with Sulfo-SANPAH solution (Thermo) for 5 min under ultraviolet light. The substrates were then washed initially with 10 mM HEPES solution in deionized water, and then three times with sterile PBS. The hydrogels were then incubated with type I collagen solution (10 μg/ml) at 4°C for 12 h [56,57]. After protein conjugation, dishes were washed three times with sterile PBS, and MEF 3T3 cells were seeded at a concentration that allowed single-cell observation and incubated in a cell incubator for the next 12 h. Cells were seeded in DMEM Low Glucose medium without Phenol Red dye (Bio West).

In each experiment, the local temperature was measured every 10 minutes. Each time point consisted of three separate measurements: snapshot in transmitted light mode to locate the cell (here DIC contrast), time-lapse ODMR signal collection, and acquisition of Z-Stack images of TFM beads. For most of the experiment time, the microwave antenna was placed in the standby position approximately 3 mm above the substrate and moved to a distance of ~250μm only for the registration of the ODMR signal. The Z-Stack (short series of images obtained in different focus positions) of the TFM beads was performed because of the lack of focus stabilization in such a custom experiment. The collection of several slices allowed us to select properly focused TFM images for the calculation of cellular traction.

The heating experiment started at a low temperature (28.4°C), which was maintained for seven consecutive time steps. Therefore, the temperature gradually increased to reach a difference of 1°C between each time step. After reaching ~37°C, the temperature was maintained at the same level until the end of the experiment. Heating was performed using an on-stage mini-incubator as well as a large incubation chamber, which could be heated to the defined temperature. The cooling experiment started at 37°C which was maintained for seven consecutive time steps. However, neither incubator was equipped with cooling equipment. Therefore, the temperature of the microscope environment with the observed sample was decreased by turning off the heating of both incubators and placing refrigerated ice packs inside a large incubation chamber. The Ice packs were added several times to decrease the temperature by ~1°C for every 10 min, and the cooling process was verified with a thermocouple that was placed inside the Petri dish. A final temperature of approximately 28°C was maintained until the end of the experiment.

TFM data processing was performed using custom software provided by Prof. Xavier Trepat from the Integrative Tissue and Cell Dynamics group at the Institute for Bioengineering of Catalonia in Barcelona, Spain.

**Literature**

1. B. Alberts, R. Heald, A. Johnson, D. Morgan, M. Raff, K. Roberts, and P. Walter, "Cell Signaling," in *Molecular Biology of the Cell, Seventh Edition* (W. W. Norton & Company, 2022), pp. 873–948.

2. M. Bielfeldt, H. Rebl, K. Peters, K. Sridharan, S. Staehlke, and J. B. Nebe, "Sensing of Physical Factors by Cells: Electric Field, Mechanical Forces, Physical Plasma and Light—Importance for Tissue Regeneration," Biomedical Materials & Devices (2022).

3. B. M. Baker and C. S. Chen, "Deconstructing the third dimension – how 3D culture microenvironments alter cellular cues," J Cell Sci **125**, 3015–3024 (2012).

4. Y. Zhang and P. Habibovic, "Delivering Mechanical Stimulation to Cells: State of the Art in Materials and Devices Design," Advanced Materials **34**, 2110267 (2022).

5. K. H. Vining and D. J. Mooney, "Mechanical forces direct stem cell behaviour in development and regeneration," Nat Rev Mol Cell Biol **18**, 728–742 (2017).

6. M. Dembo and Y. L. Wang, "Stresses at the cell-to-substrate interface during locomotion of fibroblasts," Biophys J **76**, 2307–2316 (1999).

7. J. P. Butler, I. M. Tolić-Nørrelykke, B. Fabry, and J. J. Fredberg, "Traction fields, moments, and strain energy that cells exert on their surroundings," American Journal of Physiology-Cell Physiology **282**, C595–C605 (2002).

8. J. A. Mulligan, F. Bordeleau, C. A. Reinhart-King, and S. G. Adie, "Traction Force Microscopy for Noninvasive Imaging of Cell Forces," in (2018), pp. 319–349.

9. A. Zancla, P. Mozetic, M. Orsini, G. Forte, and A. Rainer, "A primer to traction force microscopy," Journal of Biological Chemistry **298**, 101867 (2022).

10. M. Ghibaudo, A. Saez, L. Trichet, A. Xayaphoummine, J. Browaeys, P. Silberzan, A. Buguin, and B. Ladoux, "Traction forces and rigidity sensing regulate cell functions," Soft Matter **4**, 1836–1843 (2008).

11. M. Gómez-González, E. Latorre, M. Arroyo, and X. Trepat, "Measuring mechanical stress in living tissues," Nature Reviews Physics **2**, 300–317 (2020).

12. A. Brugués, E. Anon, V. Conte, J. H. Veldhuis, M. Gupta, J. Colombelli, J. J. Muñoz, G. W. Brodland, B. Ladoux, and X. Trepat, "Forces driving epithelial wound healing," Nat Phys **10**, 683–690 (2014).

13. C. M. Kraning-Rush, J. P. Califano, and C. A. Reinhart-King, "Cellular Traction Stresses Increase with Increasing Metastatic Potential," PLoS One **7**, e32572 (2012).

14. Y. Zhang, X. Shi, T. Zhao, C. Huang, Q. Wei, X. Tang, L. C. Santy, M. T. A. Saif, and S. Zhang, "A traction force threshold signifies metastatic phenotypic change in multicellular epithelia," Soft Matter **15**, 7203–7210 (2019).

15. A. A. Romanovsky, "The thermoregulation system and how it works," in (2018), pp. 3–43.

16. C. W. Meyer, Y. Ootsuka, and A. A. Romanovsky, "Body Temperature Measurements for Metabolic Phenotyping in Mice," Front Physiol **8**, (2017).

17. K. Metzger, D. Dannenberger, A. Tuchscherer, S. Ponsuksili, and C. Kalbe, "Effects of temperature on proliferation of myoblasts from donor piglets with different thermoregulatory maturities," BMC Mol Cell Biol **22**, (2021).

18. A. M. Gorbach, J. D. Heiss, L. Kopylev, and E. H. Oldfield, "Intraoperative infrared imaging of brain tumors," J Neurosurg **101**, 960–969 (2004).

19. I. S. Singh and J. D. Hasday, "Fever, hyperthermia and the heat shock response," International Journal of Hyperthermia **29**, 423–435 (2013).

20. H. Yang, C. Tu, Z. Jia, Q. Meng, J. Zhang, J. Wang, Y. Zhao, C. Zhu, and F. Bao, "Dynamic Characterization of Thermocouples under Double-Pulse Laser-Induced Thermal Excitation," Sensors **23**, 2367 (2023).

21. C. Wang, R. Xu, W. Tian, X. Jiang, Z. Cui, M. Wang, H. Sun, K. Fang, and N. Gu, "Determining intracellular temperature at single-cell level by a novel thermocouple method," Cell Res **21**, 1517–1519 (2011).

22. R. Shrestha, T.-Y. Choi, W. Chang, and D. Kim, "A High-Precision Micropipette Sensor for Cellular-Level Real-Time Thermal Characterization," Sensors **11**, 8826–8835 (2011).

23. S. Herth, M. Giesguth, W. Wedel, G. Reiss, and K.-J. Dietz, "Thermomicrocapillaries as temperature biosensors in single cells," Appl Phys Lett **102**, (2013).

24. K. Okabe, R. Sakaguchi, B. Shi, and S. Kiyonaka, "Intracellular thermometry with fluorescent sensors for thermal biology.," Pflugers Arch **470**, 717–731 (2018).

25. L. M. Maestro, E. M. Rodríguez, F. S. Rodríguez, M. C. I. la Cruz, A. Juarranz, R. Naccache, F. Vetrone, D. Jaque, J. A. Capobianco, and J. G. Solé, "CdSe Quantum Dots for Two-Photon Fluorescence Thermal Imaging," Nano Lett **10**, 5109–5115 (2010).

26. S. Arai, S.-C. Lee, D. Zhai, M. Suzuki, and Y. T. Chang, "A Molecular Fluorescent Probe for Targeted Visualization of Temperature at the Endoplasmic Reticulum," Sci Rep **4**, 6701 (2014).

27. G. Ke, C. Wang, Y. Ge, N. Zheng, Z. Zhu, and C. J. Yang, "L-DNA Molecular Beacon: A Safe, Stable, and Accurate Intracellular Nano-thermometer for Temperature Sensing in Living Cells," J Am Chem Soc **134**, 18908–18911 (2012).

28. J. Qiao, C. Chen, L. Qi, M. Liu, P. Dong, Q. Jiang, X. Yang, X. Mu, and L. Mao, "Intracellular temperature sensing by a ratiometric fluorescent polymer thermometer," J. Mater. Chem. B **2**, 7544–7550 (2014).

29. K. Okabe, N. Inada, C. Gota, Y. Harada, T. Funatsu, and S. Uchiyama, "Intracellular temperature mapping with a fluorescent polymeric thermometer and fluorescence lifetime imaging microscopy," Nat Commun **3**, 705 (2012).

30. S. Choi, V. N. Agafonov, V. A. Davydov, and T. Plakhotnik, "Ultrasensitive All-Optical Thermometry Using Nanodiamonds with a High Concentration of Silicon-Vacancy Centers and Multiparametric Data Analysis," ACS Photonics **6**, 1387–1392 (2019).

31. W. W. W. Hsiao, Y. Y. Hui, P. C. Tsai, and H. C. Chang, "Fluorescent Nanodiamond: A Versatile Tool for Long-Term Cell Tracking, Super-Resolution Imaging, and Nanoscale Temperature Sensing," Acc Chem Res **49**, 400–407 (2016).

32. G. Balasubramanian, I. Y. Chan, R. Kolesov, M. Al-Hmoud, J. Tisler, C. Shin, C. Kim, A. Wojcik, P. R. Hemmer, A. Krueger, T. Hanke, A. Leitenstorfer, R. Bratschitsch, F. Jelezko, and J. Wrachtrup, "Nanoscale imaging magnetometry with diamond spins under ambient conditions," Nature **455**, 648–651 (2008).

33. C. Y. Fang, V. Vaijayanthimala, C. A. Cheng, S. H. Yeh, C. F. Chang, C. L. Li, and H. C. Chang, "The exocytosis of fluorescent nanodiamond and its use as a long-term cell tracker," Small **7**, 3363–3370 (2011).

34. T. Genjo, S. Sotoma, R. Tanabe, R. Igarashi, and M. Shirakawa, "A Nanodiamond-peptide Bioconjugate for Fluorescence and ODMR Microscopy of a Single Actin Filament," Analytical Sciences **32**, 1165–1170 (2016).

35. D. A. Simpson, E. Morrisroe, J. M. McCoey, A. H. Lombard, D. C. Mendis, F. Treussart, L. T. Hall, S. Petrou, and L. C. L. Hollenberg, "Non-Neurotoxic Nanodiamond Probes for Intraneuronal Temperature Mapping," ACS Nano **11**, 12077–12086 (2017).

36. M. W. Doherty, N. B. Manson, P. Delaney, F. Jelezko, J. Wrachtrup, and L. C. L. Hollenberg, "The nitrogen-vacancy colour centre in diamond," Phys Rep **528**, 1–45 (2013).

37. F. Dolde, H. Fedder, M. W. Doherty, T. Nöbauer, F. Rempp, G. Balasubramanian, T. Wolf, F. Reinhard, L. C. L. Hollenberg, F. Jelezko, and J. Wrachtrup, "Electric-field sensing using single diamond spins," Nat Phys **7**, 459–463 (2011).

38. J. F. Barry, M. J. Turner, J. M. Schloss, D. R. Glenn, Y. Song, M. D. Lukin, H. Park, and R. L. Walsworth, "Optical magnetic detection of single-neuron action potentials using quantum defects in diamond," Proceedings of the National Academy of Sciences **113**, 14133–14138 (2016).

39. D. Le Sage, K. Arai, D. R. Glenn, S. J. DeVience, L. M. Pham, L. Rahn-Lee, M. D. Lukin, A. Yacoby, A. Komeili, and R. L. Walsworth, "Optical magnetic imaging of living cells," Nature **496**, 486–489 (2013).

40. A. M. Wojciechowski, P. Nakonieczna, M. Mrózek, K. Sycz, A. Kruk, M. Ficek, M. Głowacki, R. Bogdanowicz, and W. Gawlik, "Optical magnetometry based on nanodiamonds with nitrogen-vacancy color centers," Materials **12**, (2019).

41. V. M. Acosta, E. Bauch, M. P. Ledbetter, A. Waxman, L.-S. Bouchard, and D. Budker, "Temperature Dependence of the Nitrogen-Vacancy Magnetic Resonance in Diamond," Phys Rev Lett **104**, 070801 (2010).

42. G. Kucsko, P. C. Maurer, N. Y. Yao, M. Kubo, H. J. Noh, P. K. Lo, H. Park, and M. D. Lukin, "Nanometre-scale thermometry in a living cell," Nature **500**, 54–58 (2013).

43. M. Alkahtani, L. Jiang, R. Brick, P. Hemmer, and M. Scully, "Nanometer-scale luminescent thermometry in bovine embryos," Opt Lett **42**, 4812 (2017).

44. P.-C. Tsai, C. P. Epperla, J.-S. Huang, O. Y. Chen, C.-C. Wu, and H.-C. Chang, "Measuring Nanoscale Thermostability of Cell Membranes with Single Gold-Diamond Nanohybrids," Angewandte Chemie International Edition **56**, 3025–3030 (2017).

45. H. Yukawa, M. Fujiwara, K. Kobayashi, Y. Kumon, K. Miyaji, Y. Nishimura, K. Oshimi, Y. Umehara, Y. Teki, T. Iwasaki, M. Hatano, H. Hashimoto, and Y. Baba, "A quantum thermometric sensing and analysis system using fluorescent nanodiamonds for the evaluation of living stem cell functions according to intracellular temperature," Nanoscale Adv **2**, 1859–1868 (2020).

46. M. Fujiwara, S. Sun, A. Dohms, Y. Nishimura, K. Suto, Y. Takezawa, K. Oshimi, L. Zhao, N. Sadzak, Y. Umehara, Y. Teki, N. Komatsu, O. Benson, Y. Shikano, and E. Kage-Nakadai, *Real-Time Nanodiamond Thermometry Probing in Vivo Thermogenic Responses* (2020), Vol. 6.

47. H. An, Z. Yin, C. Mitchell, A. Semnani, A. R. Hajrasouliha, and M. Hosseini, "Nanodiamond ensemble-based temperature measurement in living cells and its limitations," Meas Sci Technol **32**, 015701 (2021).

48. S. R. Hemelaar, A. Nagl, F. Bigot, M. M. Rodríguez-García, M. P. de Vries, M. Chipaux, and R. Schirhagl, "The interaction of fluorescent nanodiamond probes with cellular media.," Mikrochim Acta **184**, 1001–1009 (2017).

49. X. Serra-Picamal, V. Conte, R. Vincent, E. Anon, D. T. Tambe, E. Bazellieres, J. P. Butler, J. J. Fredberg, and X. Trepat, "Mechanical waves during tissue expansion," Nat Phys **8**, 628–634 (2012).

50. M. J. Dalby, N. Gadegaard, R. Tare, A. Andar, M. O. Riehle, P. Herzyk, C. D. W. Wilkinson, and R. O. C. Oreffo, "The control of human mesenchymal cell differentiation using nanoscale symmetry and disorder," Nat Mater **6**, 997–1003 (2007).

51. R. J. McMurray, N. Gadegaard, P. M. Tsimbouri, K. V. Burgess, L. E. McNamara, R. Tare, K. Murawski, E. Kingham, R. O. C. Oreffo, and M. J. Dalby, "Nanoscale surfaces for the long-term maintenance of mesenchymal stem cell phenotype and multipotency," Nat Mater **10**, 637–644 (2011).

52. C. Allan, A. Ker, C. A. Smith, P. M. Tsimbouri, J. Borsoi, S. O'Neill, N. Gadegaard, M. J. Dalby, and R. M. Dominic Meek, "Osteoblast response to disordered nanotopography," J Tissue Eng **9**, (2018).

53. M. Chipaux, K. J. van der Laan, S. R. Hemelaar, M. Hasani, T. Zheng, and R. Schirhagl, "Nanodiamonds and Their Applications in Cells," Small **14**, 1704263 (2018).

54. N. Raval, R. Maheshwari, D. Kalyane, S. R. Youngren-Ortiz, M. B. Chougule, and R. K. Tekade, "Importance of Physicochemical Characterization of Nanoparticles in Pharmaceutical Product Development," in *Basic Fundamentals of Drug Delivery* (Elsevier, 2019), pp. 369–400.

55. T. H. Besseling, J. Jose, and A. van Blaaderen, "Methods to calibrate and scale axial distances in confocal microscopy as a function of refractive index," J Microsc **257**, 142–150 (2015).

56. Y.-L. Wang and R. J. Pelham, "[39] Preparation of a flexible, porous polyacrylamide substrate for mechanical studies of cultured cells," in *J* (1998), Vol. 68, pp. 489–496.

57. K. A. Beningo, C.-M. Lo, and Y.-L. Wang, "Flexible polyacrylamide substrata for the analysis of mechanical interactions at cell-substratum adhesions," in *Methods in Cell Biology.* (2002), Vol. 69, pp. 325–339.

## Supplementary Information – ODMR temperature calibration

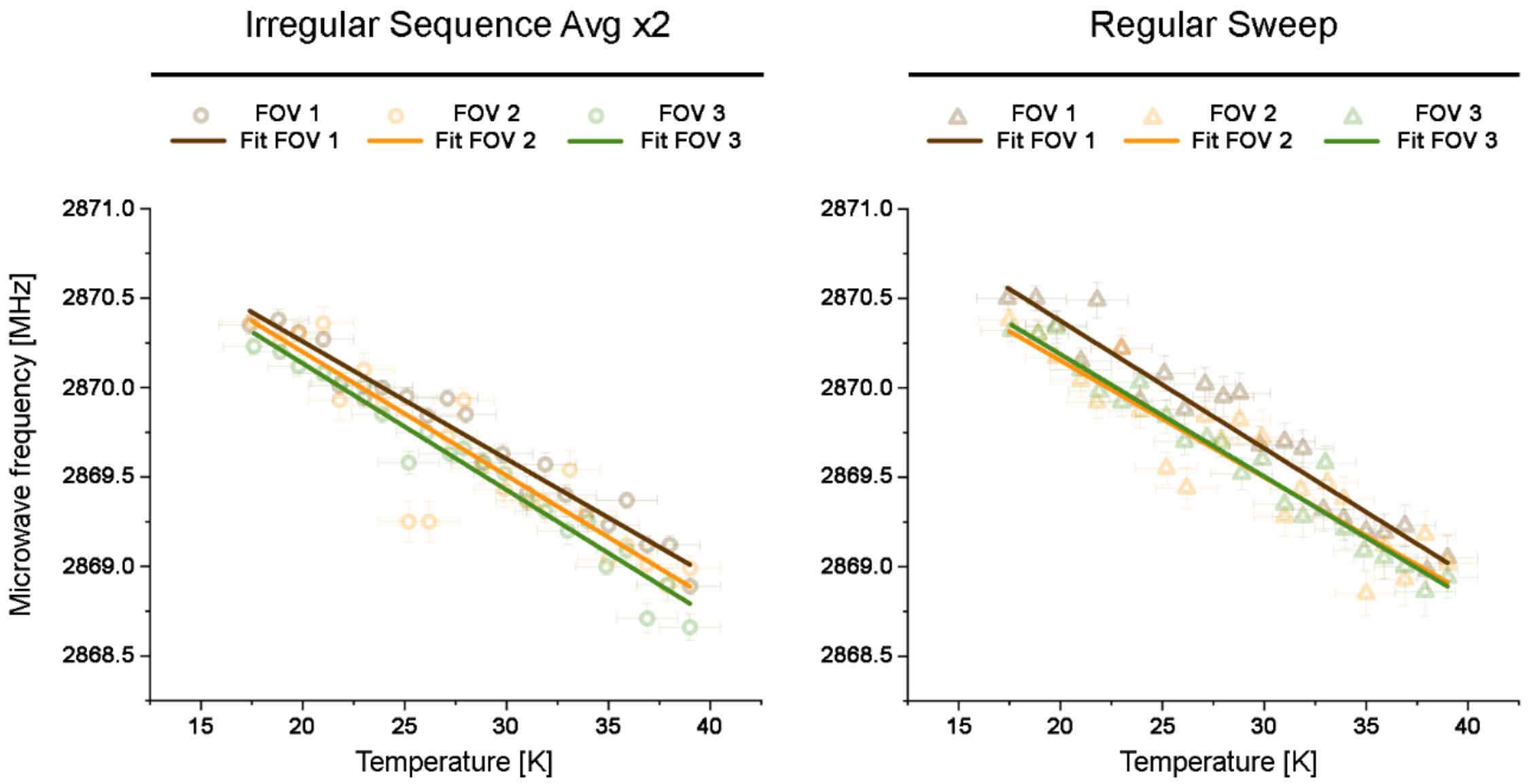

Supplementary Figure S1. Temperature calibration of ODMR-TFM substrate. Left plot shows the three experimental series for irregular sequence (t = 2.5 s) while the right plot shows the ODMR calibration for regular sweep (t = 5.1 s).

Supplementary Table S1. Slopes of linear functions fitted to the temperature calibration data

| | Irregular Seq avg x2 | | Regular Sweep | |
|---|---|---|---|---|
| | Slope [kHz/K] | $R^2$ | Slope [kHz/K] | $R^2$ |
| Position 1 | –65.6 ± 3.0 | 0,96040 | – 71.2 ± 3.8 | 0,94498 |
| Position 2 | –69.0 ± 5.3 | 0,89545 | – 65.2 ± 5.6 | 0,87062 |
| Position 3 | – 70.7 ± 2.7 | 0,97225 | – 68.3 ± 3.2 | 0,95796 |
| Averaged | – 68.4 ± 3.6 | | – 68.2 ± 4.2 | --- |